\documentclass[12pt]{iopart}
\usepackage{amsmath,amssymb,amsfonts}
\usepackage{graphicx,epsfig}
\usepackage{caption}
\usepackage{xcolor}
\usepackage{times}
\usepackage{diagbox}

\begin{document}

\title[Missing levels in intermediate spectra]{Missing levels in intermediate spectra}

\author{Mar\'{\i}a Hita-P\'{e}rez}
\address{Institute of Fundamental Physics, IFF-CSIC, Serrano 113b, Madrid, E-28006, Spain}
\ead{hitaperezmaria@gmail.com}

\author{Laura Mu\~noz}
\address{Grupo de F\'{\i}sica Nuclear, Departamento de Estructura de la Materia, F\'{\i}sica T\'ermica y Electr\'onica, and IPARCOS, Universidad Complutense de Madrid, CEI Moncloa, Madrid, E-28040, Spain}
\ead{lmunoz@ucm.es}

\author{Rafael A. Molina}
\address{Instituto de Estructura de la Materia, IEM-CSIC, Serrano 123, Madrid, E-28006, Spain}
\ead{rafael.molina@csic.es}






\begin{abstract}
We derive an expression for the nearest-neighbor spacing distribution $P(s)$ of the energy levels of quantum systems with intermediate dynamics between regularity and chaos and missing levels due to random experimental errors. The expression is based on the Brody distribution, the most widely used for fitting mixed spectra as a function of one parameter. 
By using Monte Carlo simulations of intermediate spectra based on the $\beta$-Hermite ensemble of Random Matrix Theory, we evaluate the quality of the formula 
and its suitability for fitting purposes.
Estimations of the Brody parameter and the fraction of missing levels 
can be obtained by a least-square two-parameter fitting of the experimental $P(s)$. The results should be important to distinguish the origins of deviations from RMT in experimental spectra.
\end{abstract}


\maketitle

\section{Introduction}\label{sec:intro}

Statistical analysis of spectra is a very important tool for understanding the dynamics of quantum and wave systems \cite{Mehta_Book,Stockmann_Book,Haake_Book}. Quantifying the level repulsion, for example, it is possible to study the transition between integrable and chaotic quantum or wave systems \cite{Brody1973,Brody1981,Izrailev1989,Izrailev1990,Prosen1993,Blecken1994,Molina2000,AbulMagd2008,Weidenmuller2009,Frisch2014,Munoz2017}, between localized and extended states in disordered systems \cite{Evers2008,Molina2016,Benito2017,Suntajs2021} and between ergodic and many-body localized phases in many body strongly correlated systems \cite{Oganesyan2007,Huse2014,Luitz2015,Servyn2016,Corps2021}. This relationship is based on the identification of complex wave or quantum spectra with Random Matrix Theory (RMT) \cite {Bohigas1984}. These ideas initially came from nuclear physics but are now backed up with a semiclassical justification and a great body of numerical evidence behind them \cite{Haake_Book,Heusler2007}.  Experimental studies of the spectral statistics of different quantum and wave systems from this perspective are numerous \cite{Stockmann_Book,Zimmermann1988,Stoeckmann1990,Dietz2006,Frisch2014}. However, there are intrinsic limitations that have prevented a more wide use of these tools in experimental spectra. For a meaningful statistical analysis one needs complete sequences with no missing levels, 
no mixing of symmetries and long enough to have sufficient statistics. When the spectrum to be analyzed does not fulfill these conditions the reliability of the statistical analysis can be compromised, leading to incorrect conclusions, as explained next. 

In chaotic systems with time-reversal symmetry the appropriate RMT ensemble is the GOE (Gaussian Orthogonal Ensemble)
whereas regular systems show spectral fluctuations described by Poisson statistics, corresponding to uncorrelated spectra \cite{BerryTabor}. Thus a transition from chaos to regularity manifests in the spectral fluctuations as a transition from GOE to Poisson statistics. But this kind of transition can also be induced in a GOE spectrum by the loss of correlations caused by missing levels or mixed symmetries, taking its spectral statistics also towards Poisson. Thus, when statistical analysis of a spectrum throws an intermediate result between GOE and Poisson it is quite difficult and requires a complex analysis probably taking into account different statistics \cite{Mur2015} to distinguish the actual origin of the intermediate behavior, whether it is due to a true mixed dynamics between chaos and regularity or to missing levels and/or mixed symmetries in an
actual GOE spectrum. Moreover, if it is due to both reasons it is even more difficult (when not impossible, unless one has some previous theoretical or experimental information on the dynamics of the system or the completeness of the spectrum) to find out what is the main one or estimate the weight of each, that is, estimate the degree of chaos and the number of missing levels or mixed symmetries independently.  

There has been a line of research that tries to circumvent some of these limitations and even take advantage of them in order to estimate the number of missing levels and the number of mixed symmetries in a particular experimental level sequence \cite{Brody1981,bohigaspato,Molina2007,Mulhall2011,Casal2021}. These approaches are based on RMT and the key point in order to be able to extract reliable information about missing levels or mixed symmetries is to assume that the spectral statistics coincide with the GOE (or the corresponding RMT ensemble for each symmetry class), that is, the actual experimental spectrum is chaotic. For example, analyzing the spectral statistics of chaotic nuclei it is possible to estimate how isospin symmetry is broken \cite{Mitchell1988}. The number of missing levels in experimental sequences can also be estimated using these techniques. It is possible, then, to correct the value of experimentally obtained level densities \cite{Bialous2016,Lawniczak2018,Lawniczak2017,Che2021}.

To sum up, there are tools available to determine the degree of chaos assuming that the spectrum is complete (no missing levels) and tools to estimate the number of missing levels assuming that the spectrum is chaotic (GOE). But to the best of our knowledge, there are no tools with which one can obtain both parameters at the same time: there has been no attempt to study the effect of missing levels on the spectral statistics of a mixed system between chaos and regularity. It is the purpose of this paper to fill the gap. We have been able to derive a two-parameter formula for the nearest-neighbor spacing distribution $P(s)$ that allow an independent estimation of the degree of chaos and the fraction of experimentally observed levels.
Through Monte Carlo simulations of mixed spectra based on the $\beta$-Hermite ensemble we study the accuracy of our formula and perform some tests to prove its usefulness by fitting the $P(s)$ of these simulated spectra. We prove then that our formula is useful for estimating at the same time the chaoticity and the fraction of missing levels of an experimental level sequence when this fraction is not very large. 

\section{$P(s)$ for missing levels in intermediate systems}

The two extremes (regular-Poisson and chaotic-GOE) in the nearest-neighbor spacing distribution (NNSD) are universal. However, there is not a universal transition for the intermediate dynamics. Two general behaviors in this transition can be distinguished in terms of level repulsion: fractional level repulsion and level repulsion of only a fraction of levels \cite{Berry1984,Prosen1994}. Fractional level repulsion when the NNSD behaves as $P(s) \sim s^q$ for small values of $s$ is the most common.
It can be described phenomenologically quite well by
the Brody \cite{Brody1973} and the Izrailev \cite{Izrailev1989} distributions. Both describe this kind of repulsion at small spacings but are slightly different at large spacings. They are phenomenological distributions, although the Brody distribution can be derived from a power-law ansatz for the level repulsion function \cite{Engel1998}.
We choose the Brody distribution in this paper as it is the more extensively used for fitting results in intermediate systems.
The Brody's formula is given by:
\begin{equation}\label{eq:brody}
    P_B(s)=as^q\exp (-bs^{q+1}),
\end{equation}
where $a=(q+1)b$, $b=\left[\Gamma(\frac{q+2}{q+1})\right]^{q+1}$, and $q$ is the so called Brody's parameter or mixing parameter, which allows to interpolate between Poisson distribution for $q=0$ (regularity):
\begin{equation}\label{eq:Poisson}
    P_P(s)=\exp (-s),
\end{equation}
and Wigner distribution for GOE for $q=1$ (chaos):
\begin{equation}\label{eq:GOE}
    P_W(s)=\frac{\pi}{2}s\exp \left(-\frac{\pi}{4}s^2\right).
\end{equation}

The NNSD is the first of a series of spacing distributions between neighbors of $n-$order, $p(n,s)$, $n=0$ being the NNSD. These distributions are defined after the unfolding of the spectrum where the level density is locally normalized to unity \cite{Haake_Book}. A proper unfolding is not an easy task \cite{Gomez2002} but, for our purposes, we assume that the spectrum has been properly unfolded. Moreover, for our calculations here with $\beta$-ensemble spectra the unfolding is straightforward as the level densities are known from RMT. For unfolded levels these distributions need to be normalized and centered with the following conditions:
\begin{eqnarray}
   & \int^{\infty}_0p(n,s)ds=1, \label{eq:condnorm1}\\
&    \langle s\rangle=\int^{\infty}_0sp(n,s)ds=n+1. \label{eq:condnorm2} 
\end{eqnarray}\par

When levels are missing from a sequence the higher order spacing distributions start to play a role. Using simple statistical considerations, Bohigas and Pato derived the general formula for the NNSD of a sequence with only a fraction of observed levels $f$ (the fraction of missing levels would be then $1-f$) as a function of the $p(n,s)$ distributions and assuming that the missing levels are randomly distributed \cite{bohigaspato}:
\begin{equation}\label{eq:generalpsf}
p(s)=\sum_{k=0}^{\infty}(1-f)^{k}p(k,s/f).
\end{equation}
The interpretation of this formula is easy as the prefactor $(1-f)^{k}$ for $k \geq 1$ is the probability that there are $k$ missing levels between two given levels. An observed nearest neighbor spacing (left hand side) can really be a 2nd neighbor spacing with 1 missing level in between, a 3rd neighbor spacing with 2 missing levels in between,... a $(k+1)$th neighbor spacing with $k$ missing levels in between... in the original spectrum (right hand side).
One, then, needs to use $s/f$ instead of $s$, given that the level sequence with a fraction $f$ of observed levels is unfolded to $<s>=1$, so the average spacing in the original sequence without missing levels is a factor $1/f$ higher.

\begin{figure}
    \centering
        \includegraphics[width=0.45\textwidth]{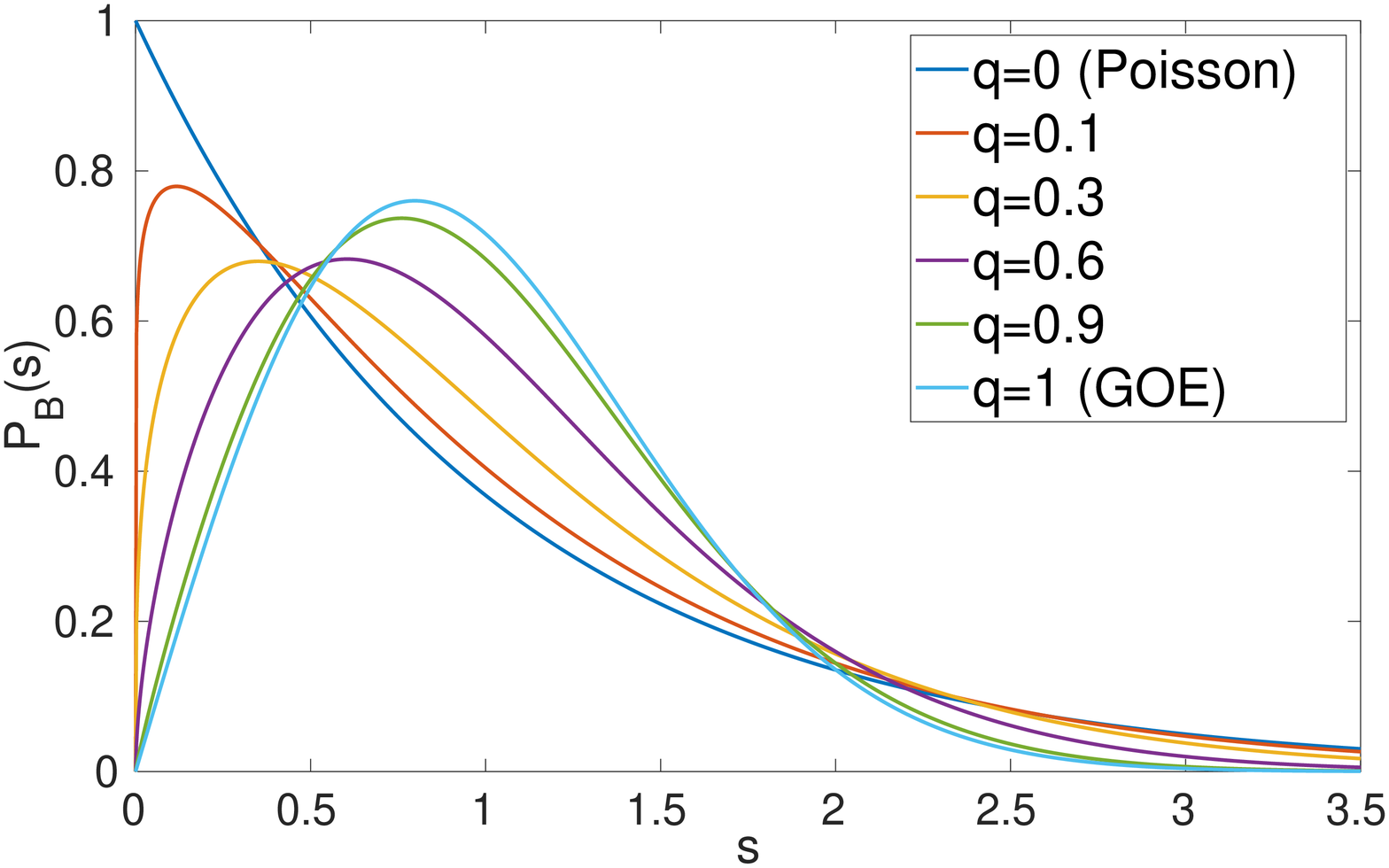}
        \includegraphics[width=0.45\textwidth]{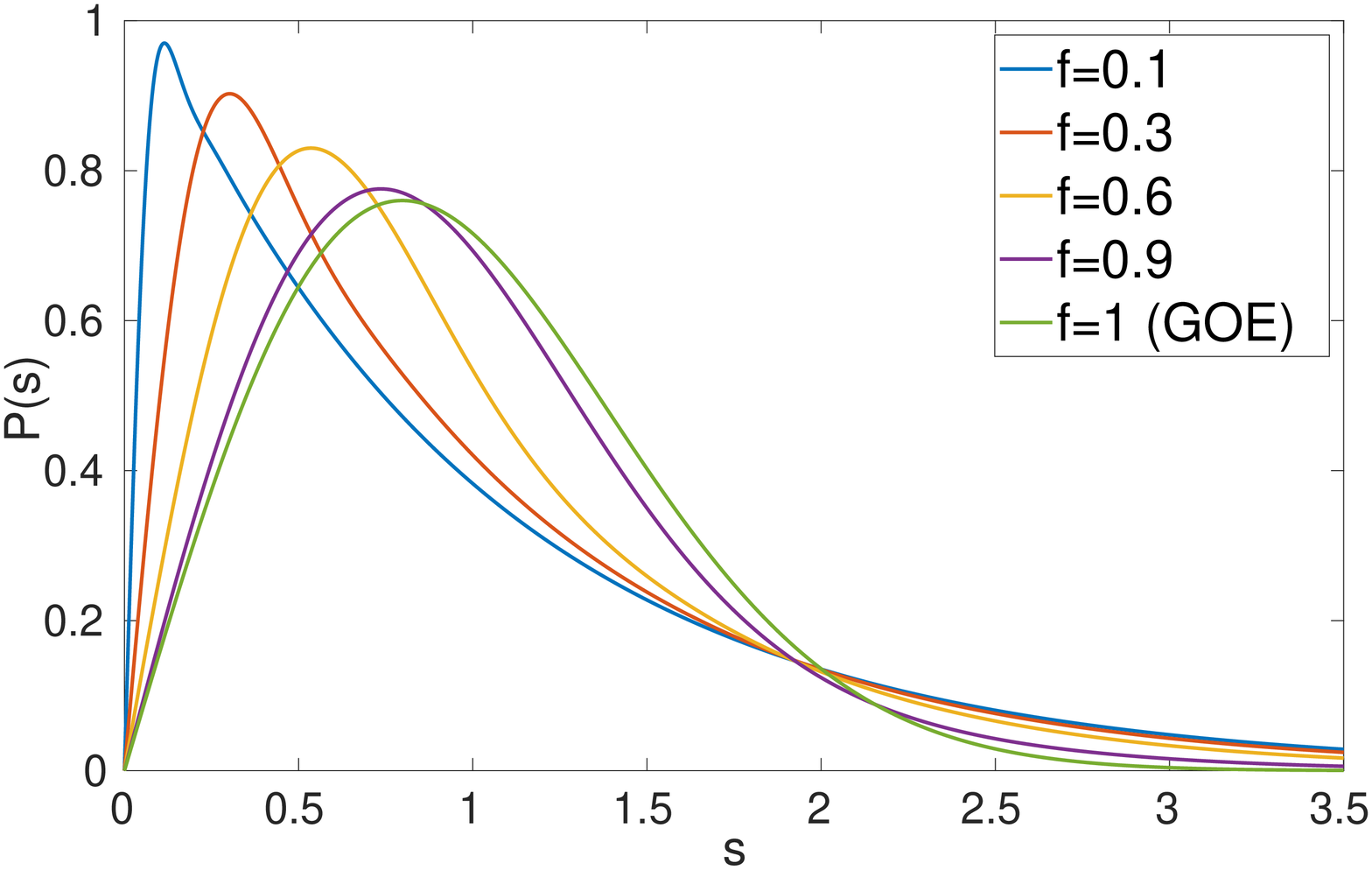}
    \caption{Comparison of the evolution of $P(s)$ from GOE to Poisson with the mixing parameter $q$ of Eq. \eqref{eq:brody} for the transition with the type of dynamics (a) and with the fraction $f$ of observed levels of Eq. \eqref{eq:generalpsf} (b).}
    \label{fig:behavior-Brody-BP}
\end{figure}

From Eq. (\ref{eq:generalpsf}) and the form of the higher-order spacing distibution for GOE one can see that the small spacing behavior of the transition from GOE to Poisson when we have fractional behavior ($P(s) \sim s^{q}$) is very different from the transition to Poisson due to missing levels where we still have $P(s) \sim s$. This can be seen in Fig. \ref{fig:behavior-Brody-BP}, where we show the evolution with the type of dynamics of Eq. \eqref{eq:brody} for different values of $q$ compared to the evolution with missing levels of Eq. \eqref{eq:generalpsf} for different values of $f$. This gives us hope that we can derive a two-parameter distribution that can distinguish the origin of the transition through a fit to experimental spectra.

Given Brody's nearest-neighbor spacing distribution \eqref{eq:brody}, Abul-Magd and Simbel \cite{AbulMagd2000} derived a general expression for the level-spacing distributions for higher order neighbors $p(n,s)$ using a statistic treatment proposed by Engel, Main and Wunner \cite{Engel1998}. They obtained the following generalization of Brody's formula:
\begin{equation}\label{eq:pnsbrody}
    p_B(n,s)=a_ns^{(n+1)q}\exp{(-b_ns^{(n+1)q+1})} 
    \int^s_0 p(n-1,x) \exp(b_nx^{(n+1)q+1})dx, 
\end{equation}
for $n\geq 1$. Here $a_n=[(n+1)q+1]b_n$ and both constants, $a_n$ and $b_n$, are determined by $p(n,s)$ normalization conditions \eqref{eq:condnorm1} and \eqref{eq:condnorm2}.

This expression presents many complications when trying to calculate high-order spacings. This is due to the fact that an exact solution for $b_n$ can only be found in the case in which the system is regular ($q=0$) where $b_n=1$, and in all the other cases it has to be calculated by solving the integrals numerically. Those calculations are too heavy for practical purposes but we have found that Gaussian approximations for $n\geq3$ work very well. This trick was already used in the original work of Bohigas and Pato for missing levels in GOE \cite{bohigaspato}. Then, we have proceed as follows.

For $n=1$, a generalization  of the Brody formula for the next-nearest-neighbor distribution is obtained by substituting equation \eqref{eq:brody} for $p_B(0,s)$ into equation \eqref{eq:pnsbrody}.
\begin{equation}\label{eq:p1sbrody}
    p_B(1,s)=aa_1\int^s_0 s^{2q}x^q\exp\left[-bx^{q+1}-b_1(s^{2q+1}-x^{2q+1})\right]dx,
\end{equation}
 where $a_1=(2q+1)b_1$. The parameter $b_1$ was parametrized by Abul-Magd and Simbel and has the form:
\begin{equation}
 b_1(q)=\frac{1}{1+2.7q+3.5q^2}.
\end{equation}\par

For $n=2$, following the same steps and substituting equation \eqref{eq:p1sbrody} into equation \eqref{eq:pnsbrody}, we find that 
\begin{eqnarray}\label{eq:p2sbrody}
    &p_B(2,s) =aa_1a_2\int^s_0 \int^x_0 s^{3q}x^{2q}y^q\exp\left[-by^{q+1} \right. \nonumber \\
    &\left. -b_1(x^{2q+1}-y^{2q+1})-b_2(s^{3q+1}-x^{3q+1})\right]dxdy.
\end{eqnarray}
Here $a_2=(3q+1)b_2$ and $b_2$ has to be obtained numerically from the normalization conditions. As Abul-Magd and Simbel did for $b_1$, we have calculated using Monte Carlo simulations the values of $b_2$ for a scan of values of $q$ ($q=0.1,0.2,\ldots 0.9,1$) and parametrized the result in the form:
\begin{equation}
 b_2(q)=\frac{1}{1+6.7q+1.3q^2+51q^3}.
\end{equation}\par

For $n\geq3$, we found that the \textit{n}th-order level-spacing distribution follows mostly a Gaussian distribution making the calculation of $b_n$ not worth the time it takes, especially taking into account that one should keep many terms to have a good approximation to the infinite sum of expression \eqref{eq:generalpsf}. In this work we have kept up to 150 terms. Thus instead of using the expression \eqref{eq:pnsbrody} to describe the $p(n,s)$ distribution, we use now a Gaussian approximation given by
\begin{equation}\label{gaussian}
p(n,s)=\frac{1}{\sqrt{2\pi}\sigma(n,q)}\exp\left[-\frac{(s-\mu)^2}{2\sigma(n,q)}\right], \qquad n\geq3,
\end{equation}
where $\mu$ is the mean of the distribution, $\mu=\langle s \rangle=n+1$, and $\sigma(n,q)$ is its standard deviation. From now on, $\sigma(n,q)$ will be calculated using spline interpolation from a battery of $\sigma(n,q)$ values previously obtained from spectra with different $q$ values ($q=0.1,0.2,\ldots 0.9,1$). 

Once we have an expression for $p(n,s)$, we are ready to propose an expression to describe the nearest-neighbor spacing  distribution, $P(s)$,  associated  with  incomplete  spectra  of  intermediate  and  chaotic systems. Substituting equations \eqref{eq:brody}, \eqref{eq:p1sbrody}, \eqref{eq:p2sbrody} and \eqref{gaussian} into equation \eqref{eq:generalpsf}, we obtain:
\begin{equation}\label{final2}
\begin{split}
    P(s,f,q) & =a\left(\frac{s}{f}\right)^q\exp\left[-b\left(\frac{s}{f}\right)^{q+1}\right]\\
    & +(1-f)aa_1\int^{\frac{s}{f}}_0 \left(\frac{s}{f}\right)^{2q}x^q\exp\left\{-bx^{q+1}-b_1\left[\left(\frac{s}{f}\right)^{2q+1}-x^{2q+1}\right]\right\}dx\\
    & +(1-f)^2aa_1a_2\int^{\frac{s}{f}}_0 \int^x_0 \left(\frac{s}{f}\right)^{3q}x^{2q}y^q\exp\Biggl\{-by^{q+1}-b_1\left[x^{2q+1}-y^{2q+1}\right]\Biggr.\\
    & \left. -b_2\left[\left(\frac{s}{f}\right)^{3q+1}-x^{3q+1}\right]\right\}dydx\\
    & +\sum_{n\geq3}\frac{(1-f)^n}{\sqrt{2\pi}\sigma(n,q)}\exp\left\{-\frac{\left[\left(\frac{s}{f}\right)-n-1\right]^2}{2\sigma(n,q)}\right\},
\end{split}
\end{equation}
where all the parameters have already been defined.

This is the central result of this work. Despite its complicated appearance it is suitable for calculations and fitting purposes, as shown in the next sections. But besides its practical interest, the two-parameter formula is interesting in its own as a first step to start to unravel the puzzle of the actual origin of intermediate behavior of fluctuations in a given spectrum. 

\section{Comparison with $\beta$-ensemble results}

The $\beta$-Hermite random matrix ensemble was proposed as a continuous generalization for all values of the repulsion parameter $\beta>0$ of the classical random matrix ensembles corresponding to integer values $\beta=1$ (GOE), 2 (GUE) and 4 (GSE) \cite{Dumitriu}. It allows a transition between the Poisson integrable results corresponding to $\beta=0$ and the chaotic GOE results $\beta=1$ with fractional level repulsion. Moreover, due to its simple form the use of the $\beta$-Hermite ensemble results in an unrivalled speed-up of numerical simulations. These features make the $\beta$-Hermite ensemble the best choice for checking our two-parameter formula for $P(s)$ with a huge amount of matrices of different sizes with a fine scan of values of $\beta$ and fractions $f$ of observed levels. The precise value of $P(s)$ for $\beta$-Hermite spectra does not follow exactly the Brody distribution but a more complicated two parameter formula, though the Brody formula is a good description except for low values of $\beta$ \cite{LeCaer2007}. However, this fact also fits our purpose as we want a practical distribution that can be used to obtain a good estimation of the fractional level repulsion and the fraction of missing levels in any type of system without worrying if it follows exactly the Brody distribution. We summarize below how the matrices belonging to the $\beta$-Hermite ensemble are constructed.

The matrices in the $\beta$-Hermite ensemble are real symmetric tridiagonal matrices whose matrix elements are constructed as:
%
\begin{equation}
H_{\beta}  =  \frac{1}{\sqrt{2}} \left( \begin{array}{ccccc}
N(0,2) &  \chi_{(N-1)\beta} & & & \\
\chi_{(N-1)\beta} & N(0,2)  & \chi_{(N-2)\beta} & & \\
& \ddots & \ddots & \ddots & \\
& & \chi_{2\beta} & N(0,2) & \chi_{\beta} \\
& &  & \chi_{\beta} & N(0,2)
\end{array} \right),
\label{eq:betaHermite}
\end{equation}
That is, the diagonal matrix elements are random variables with a Gaussian distribution $N(\mu,\sigma^2)$ of zero mean and variance $\sigma^2=2$ while the non-diagonal matrix elements come from a $\chi_\nu$ distribution with $\nu=\beta(N-n)$, $N$ being the dimension of the matrix. 
 
The level density of the $\beta$-Hermite ensemble is the semicircle law as in the case of the GOE, so the unfolding is easy in this case:
\begin{equation}
    \bar{\rho}(E)=\frac{1}{\pi\beta}\sqrt{2N\beta-E^2}, \, \lvert E \rvert < \sqrt{2N\beta}.
\end{equation}

In order to check the accuracy of the formula for $P(s,f,q)$, Eq. (\ref{final2}), we need spectra of as high dimension as possible. Thus, we have generated ensembles of 1000 spectra from $\beta$-Hermite matrices of size $N=10000$ (equivalent to have spectra of dimension $10^7$) and values of the repulsion parameter $\beta=0,0.1,0.2,\ldots,0.9,1$ where we take out a certain number of levels $N(1-f)$ in order to have a fraction $f$ of observed levels.

\begin{figure}
    \centering
    \includegraphics[width=0.3\textwidth]{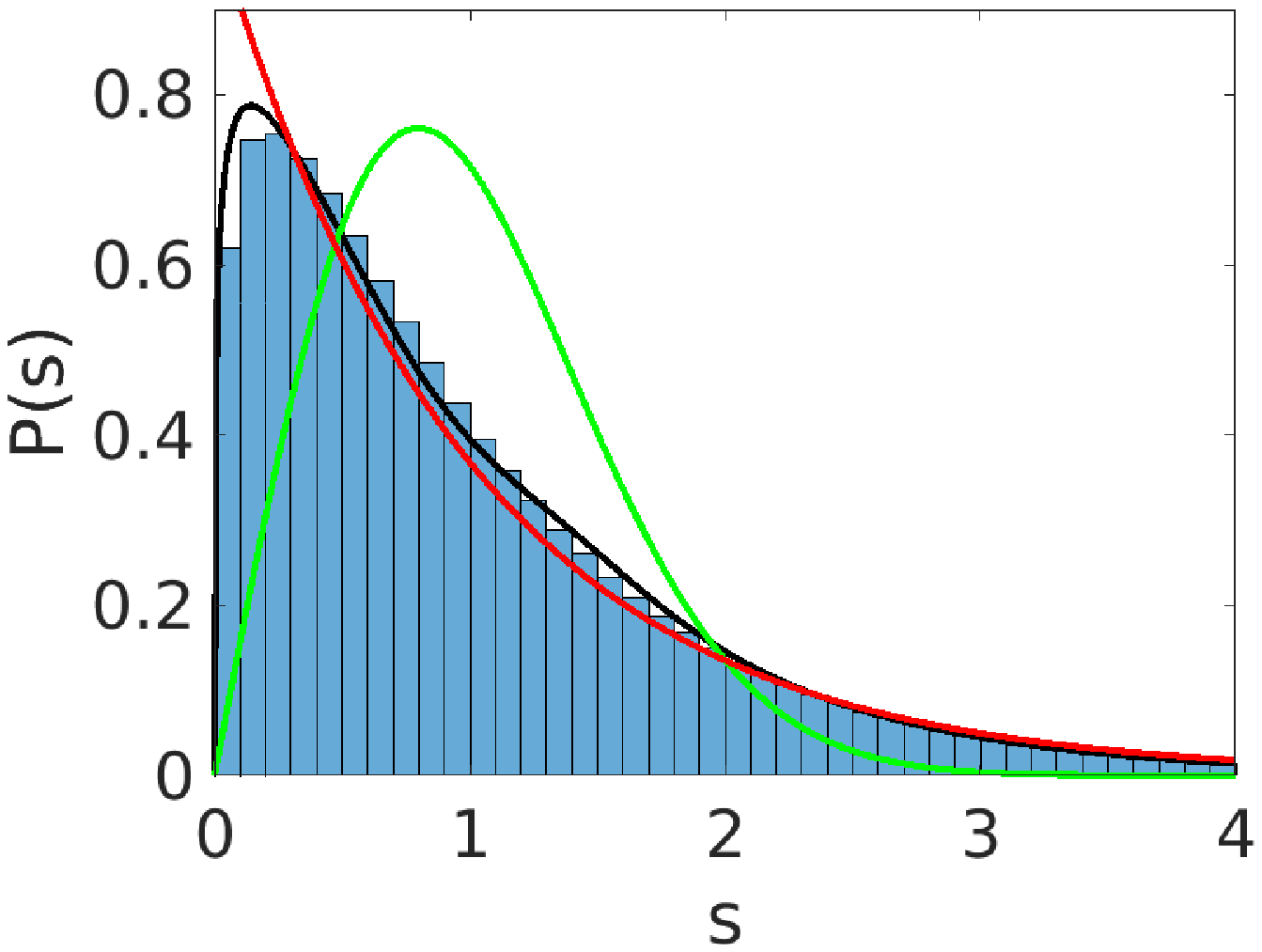}
    \includegraphics[width=0.3\textwidth]{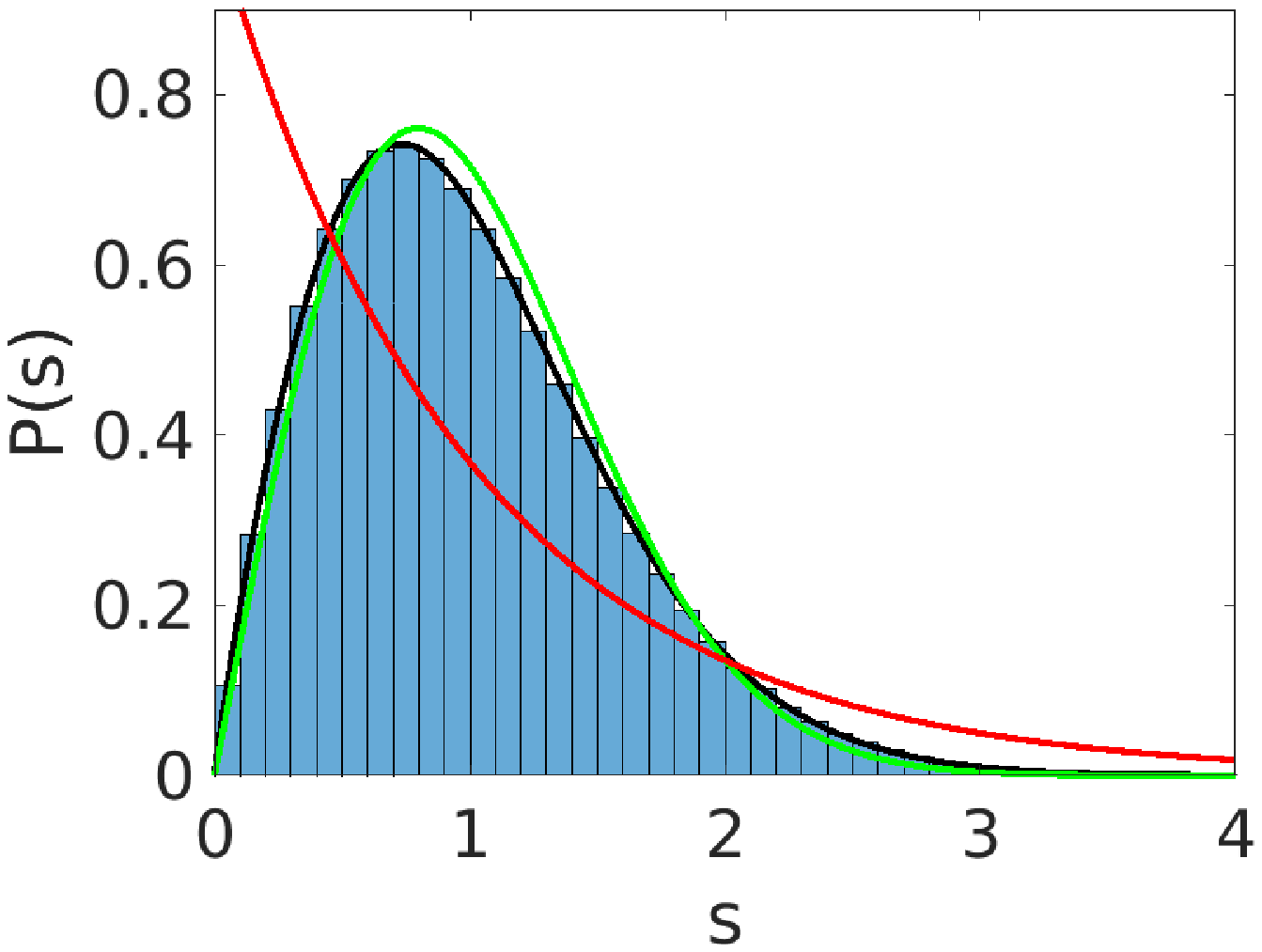}
    \includegraphics[width=0.3\textwidth]{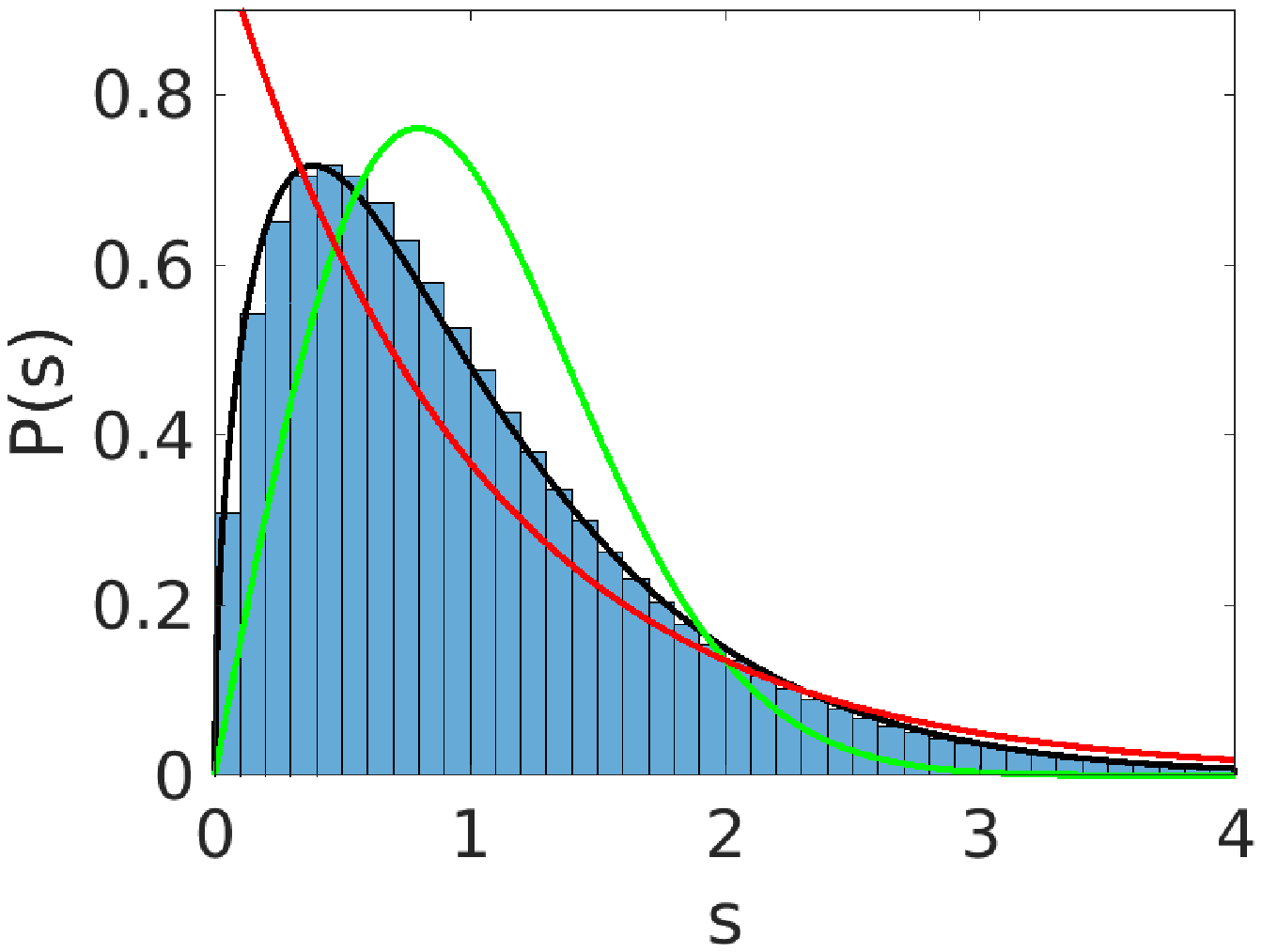}
    \includegraphics[width=0.3\textwidth]{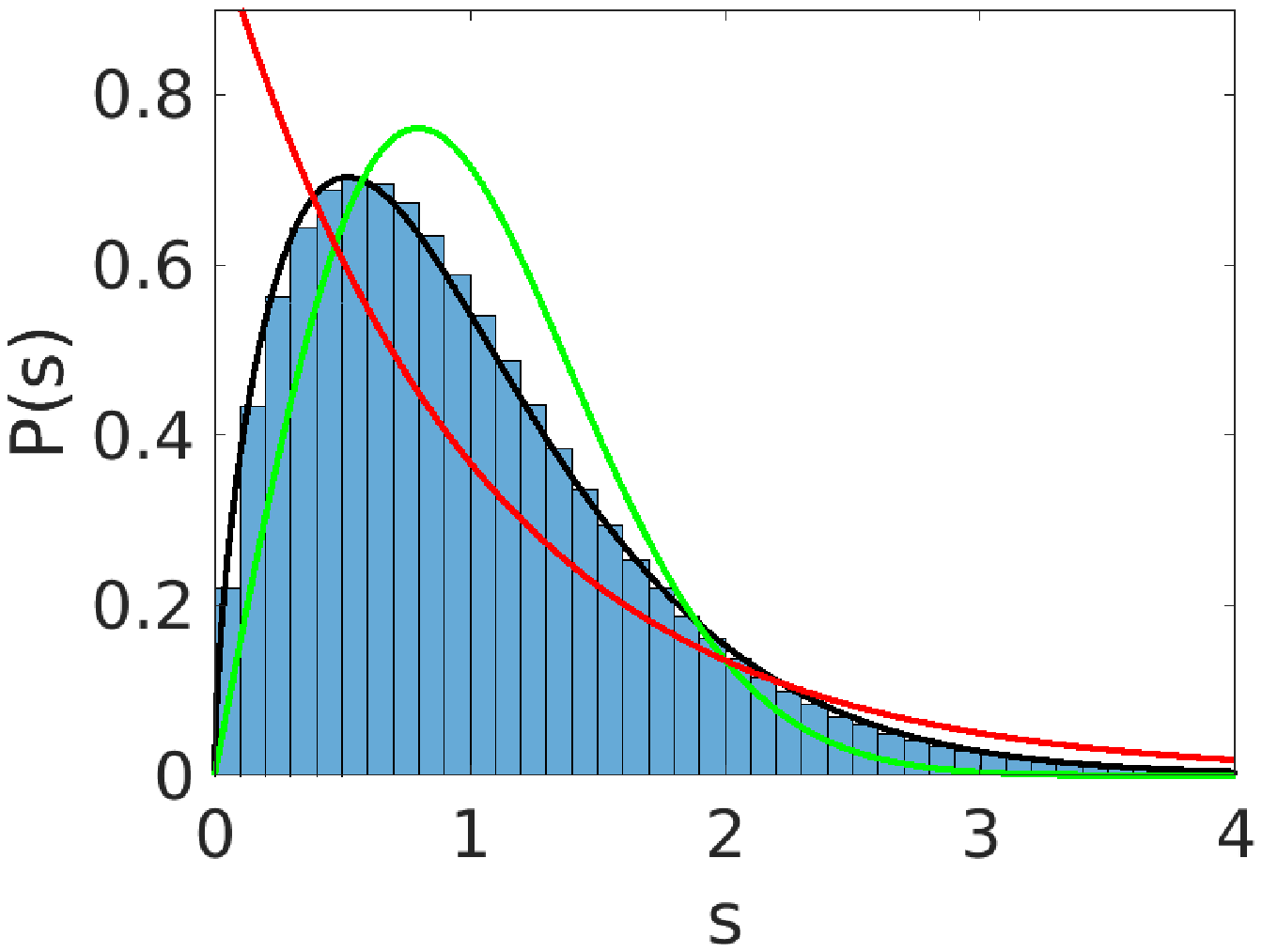}
    \includegraphics[width=0.3\textwidth]{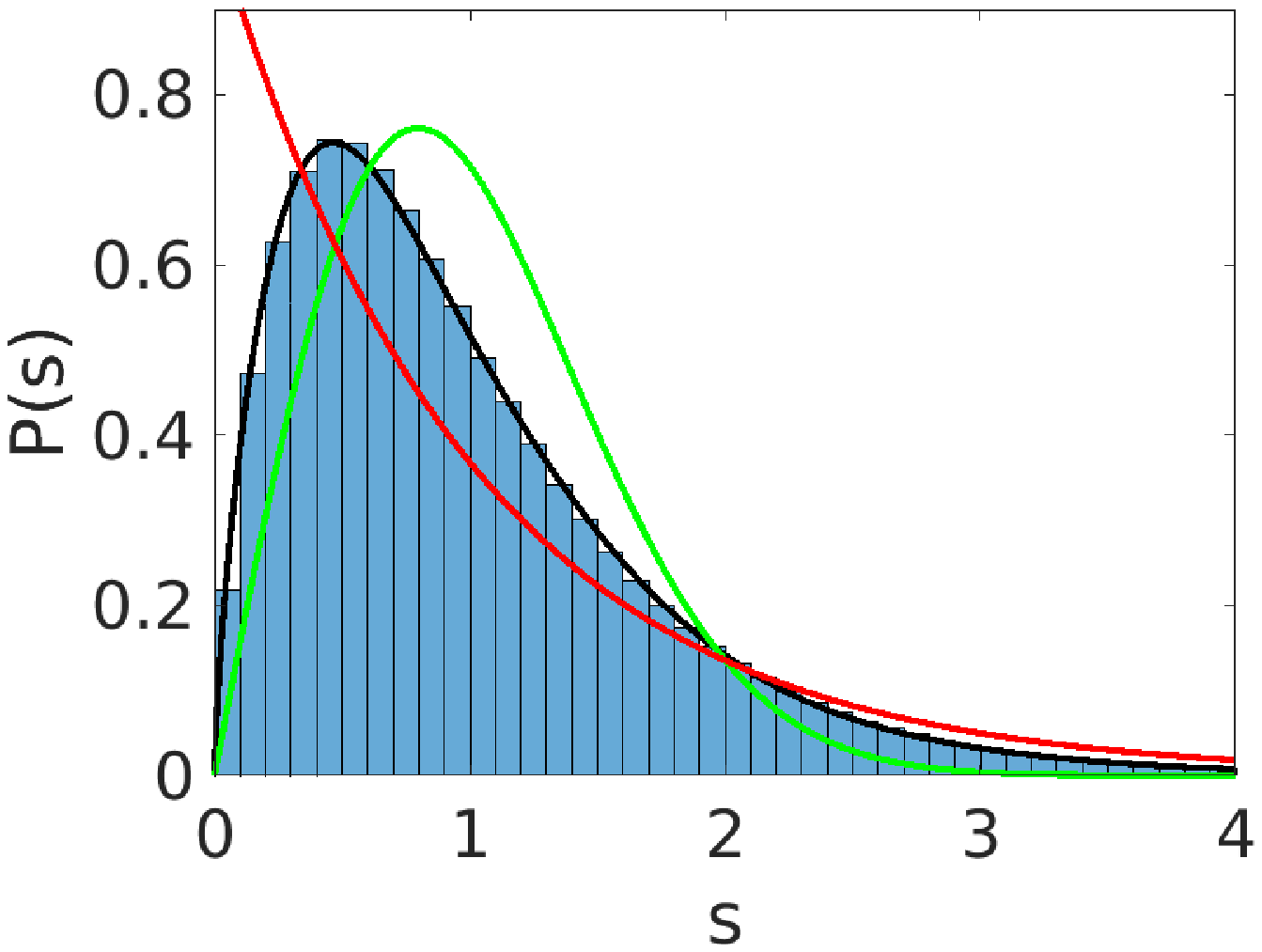}
    \includegraphics[width=0.3\textwidth]{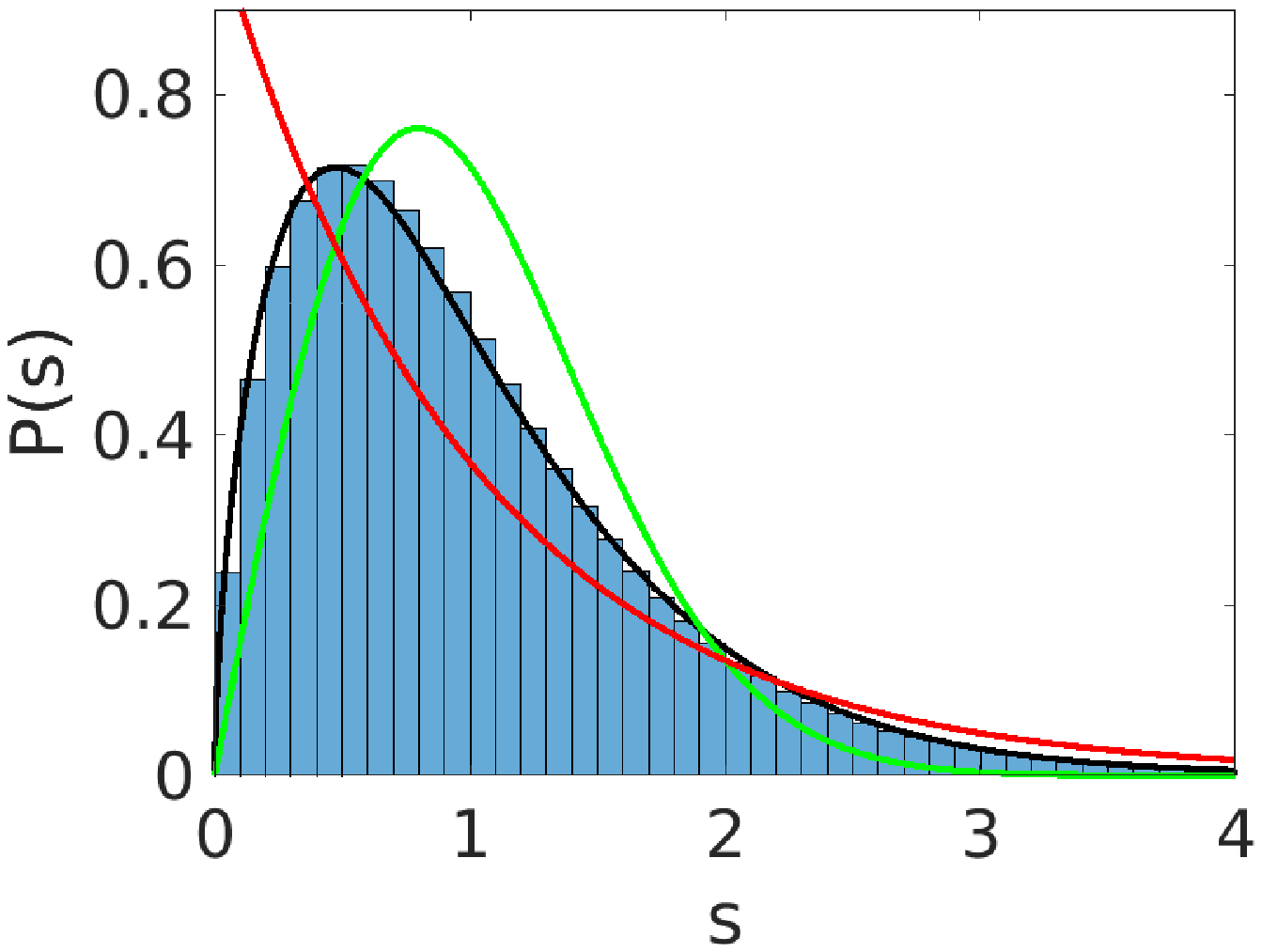}
    \caption{Comparison of Eq.\ref{final2} (black line) with ensemble averages of 1000 matrices of dimension $N=10000$ from the $\beta-$Hermite ensemble with $\beta=q$ and a fraction $f$ of observed levels (that is, the theoretical formula, not a fit). Green line is Eq. \ref{eq:GOE} (GOE) and red line is Eq. \ref{eq:Poisson} (Poisson). The values of ($f$, $q$) in each panel are: top-left (0.3, 0.2), top-middle (0.95, 0.9), top-right (0.6, 0.5), bottom-left (0.8, 0.6), bottom-middle (0.6, 0.7), bottom-right (0.7, 0.6).}
    \label{fig:comprob}
\end{figure}

In Fig. \ref{fig:comprob} we show the comparison of our formula $P(s,f,q)$ (with $q=\beta$) with the $P(s)$ obtained from these ensemble averages in two extreme cases (very low and very high values of $\beta$ and $f$) and four cases with intermediate values, which are the ones of practical interest. In Table \ref{tab:chi2} we show the values of $\chi^2$ for all the distributions we have calculated. We have found that the agreement is excellent in most cases. We make additional comments on two cases below.

First, looking at the figures in the region of very low values of $\beta$ and $f$ the $P(s)$ distributions are very close to Poisson statistics but, being the $\beta$-ensembles compatible with fractional level repulsion, the transition from $\beta > 0$ to $\beta=0$ cannot be smooth and that is why, even though the values of $\chi^2$ shown in Table \ref{tab:chi2} in this region are small, the curves are not as similar to the one for Poisson statistics as expected. In any case, we have scanned all these values of $\beta$ and $f$ for the sake of completeness but this region is of no much practical interest. No one can expect reasonable estimations with too low fractions of observed levels in too uncorrelated spectra.

Second, looking at Table \ref{tab:chi2} there is another region of values which could be strikingly higher than the rest: $f\geq0.9$, $q\leq0.3$, where $\chi^2$ is of order tenths whereas the rest of values are of order hundredths or less. However, this is also expected as this region correspond to the fitting of Brody's formula for low values of $\beta$ for a practically complete spectrum ($f\simeq1$), and, as we have mentioned before, the $P(s)$ of the $\beta$-ensembles deviate from the Brody distribution in this case \cite{LeCaer2007}.

 
In summary, in this section we made use of the $\beta$-Hermite ensemble to show that the derivation of the two-parameter formula is correct by checking its accuracy with very high dimension. In the next section we explore if the formula could be used for realistic spectra to gain information confidently about the chaoticity and the number of missing levels simultaneously by fitting data with our expression.

\begin{table}[]
\centering
\begin{tabular}{|r|r|r|r|r|r|r|r|r|}
\hline
\diagbox{
$q$}{$f$} & 0.1 & 0.3 & 0.6 & 0.7 & 0.8 & 0.9 & 0.95 & 1 \\ \hline
0.1                      & 0.021        & 0.040        & 0.052        & 0.056        & 0.063        & 0.073        & 0.078         & 0.082      \\ 
0.2                      & 0.017        & 0.035        & 0.054        & 0.063        & 0.076        & 0.091        & 0.10          & 0.11       \\ 
0.3                      & 0.013        & 0.029        & 0.048        & 0.054        & 0.066        & 0.082        & 0.092         & 0.10       \\ 
0.4                      & 0.013        & 0.022        & 0.035        & 0.041        & 0.049        & 0.061        & 0.071         & 0.083      \\ 
0.5                       & 0.013        & 0.016        & 0.023        & 0.027        & 0.031        & 0.039        & 0.050         & 0.057      \\ 
0.6                      & 0.013        & 0.0096        & 0.012        & 0.014        & 0.016        & 0.020        & 0.024         & 0.032      \\ 
0.7                       & 0.012        & 0.0062       & 0.0059       & 0.0065       & 0.0075       & 0.0076       & 0.0089        & 0.013      \\ 
0.8                       & 0.015        & 0.0079       & 0.0087       & 0.0083       & 0.0076       & 0.0054       & 0.0034        & 0.0022     \\ 
0.9                       & 0.019        & 0.018        & 0.023        & 0.022        & 0.020        & 0.015        & 0.0096         & 0.0014     \\ 
1                         & 0.030        & 0.047        & 0.054        & 0.051        & 0.045        & 0.038        & 0.029         & 0.012      \\ \hline
\end{tabular}
    \caption{Values of $\chi^2$ for the NNSD of $\beta$-ensembles of 1000 matrices of dimension $N=10000$ generated with mixing parameter $q=\beta$ and fraction of observed levels $f$ with respect to the formula of Eq. \ref{final2}.}
    \label{tab:chi2}
\end{table}

\section{Distributions of fitted results}

In this section we again take advantage of the $\beta$-ensemble in order to show that the two-parameter formula is suitable for fitting purposes and the possible limitations one should be careful with. We generate ensembles of matrices with certain values of the dimension $N$ and the parameters $\beta$ and $f$ as in the previous section, but now we perform fits of these data with our two-parameter formula and we represent in a two-dimensional histogram the distribution of the results for $q$ and $f$, so we can analyze the probability of obtaining the correct values of the parameters when analyzing a single spectrum of interest. Thus, the uncertainty in the estimation of the parameters is not only related to errors in the fit but to the variance of the parameter distribution in the proper ensemble \cite{Casal2021}. This variance increases when the matrix size is reduced.

We want to stress that we do not intend to give a general recipe on how to use the formula. We have used the $\beta$-ensemble as a simple RMT model to simulate the transition from regularity to chaos. However, the transition is not universal and can, then, be different in actual experimental spectra. So we will not present here results for a fine scan of both parameters as in the previous section, but only show some representative cases to explain important aspects to take into account and give a general recommendation. That is, we make available the formula and await for experimental results to evaluate its real applicability and usefulness when approaching each case with its particular features.

\begin{figure}
    \centering
    \includegraphics[width=\textwidth]{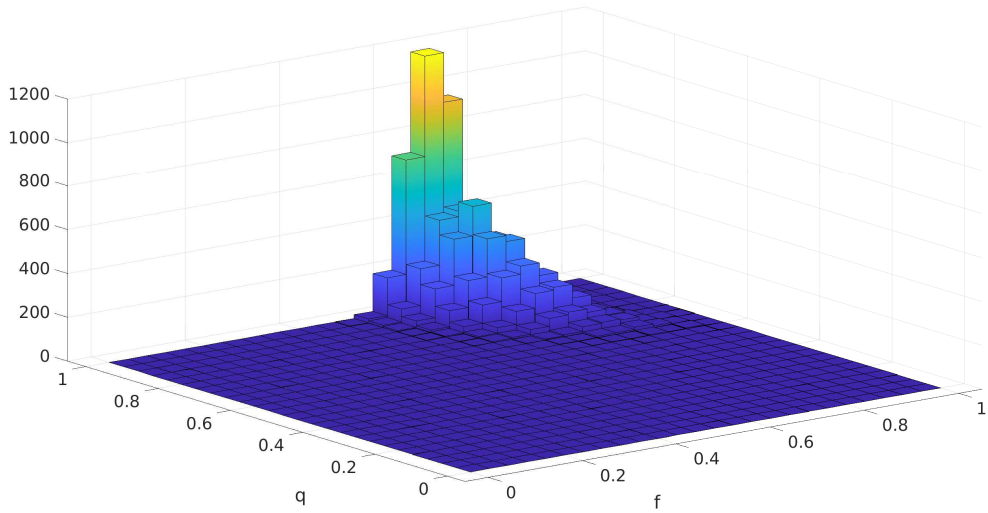}
    \caption{Histogram of the joint distribution of fitted values of the parameters ($q,f$) of an ensemble of 10000 $\beta$-Hermite matrices with $f=0.8$ and $\beta=0.9$. The matrix sizes are $N=1000$.}
    \label{fig:disp-high}
\end{figure}

For few missing levels and dynamics close to chaos (high values of $f$ and $\beta$), the spectra are still very correlated and we can obtain reliable estimations of both parameters. We show an example in Fig. \ref{fig:disp-high} where we have represented the results for the joint distribution of fitted parameters for an ensemble of 10000 spectra of dimension $N=1000$ with $f=0.8$ and $\beta=0.9$. Here we find a narrow peak of values of the parameters concentrated around the correct estimations. The ensemble variance is small in this case, so the probability of obtaining the correct values of the parameters $q$ and $f$ from a fit of the data is high.

\begin{figure}
    \centering
    \includegraphics[width=\textwidth]{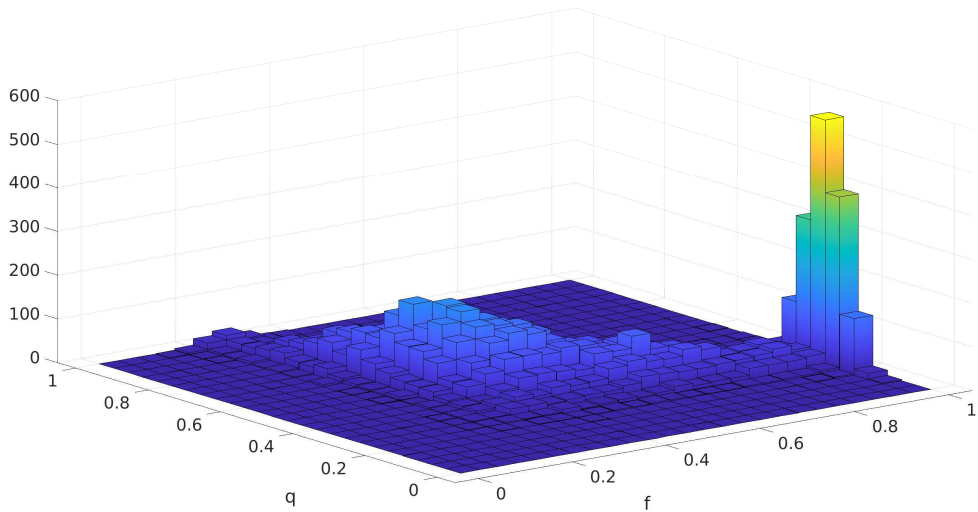}
    \caption{Histogram of the joint distribution of fitted values of the parameters ($q,f$) of an ensemble of 10000 $\beta$-Hermite matrices with $f=0.6$ and $\beta=0.3$. The matrix sizes are $N=1000$.}
    \label{fig:disp-low}
\end{figure}

However, it is obvious that when spectra are very uncorrelated (very close to Poisson statistics) little information can be gained. The loss of correlations can occur when there are many missing levels (low values of $f$) or when the dynamics is very close to regularity (low values of $\beta$). This intuition should be reflected in a very high ensemble variance of the parameters. For example, in Fig. \ref{fig:disp-low} we show the results of the distribution of the fitted parameters of the two-parameter formula for $P(s)$ in an ensemble of 10000 spectra of dimension $N=1000$ with $f=0.6$ and $\beta=0.3$. As can be seen the results for $(q,f)$ are spread over a wide region. One obtains with a high probability values around ($f  \simeq 1, q \simeq 0.3$) and, as expected, there are many different combinations of values ($q,f$) to reproduce such an uncorrelated spectra as any NNSD in which any of the two parameters is very low will be very close to Poisson.

\begin{figure}
    \centering
        \includegraphics[width=0.32\textwidth]{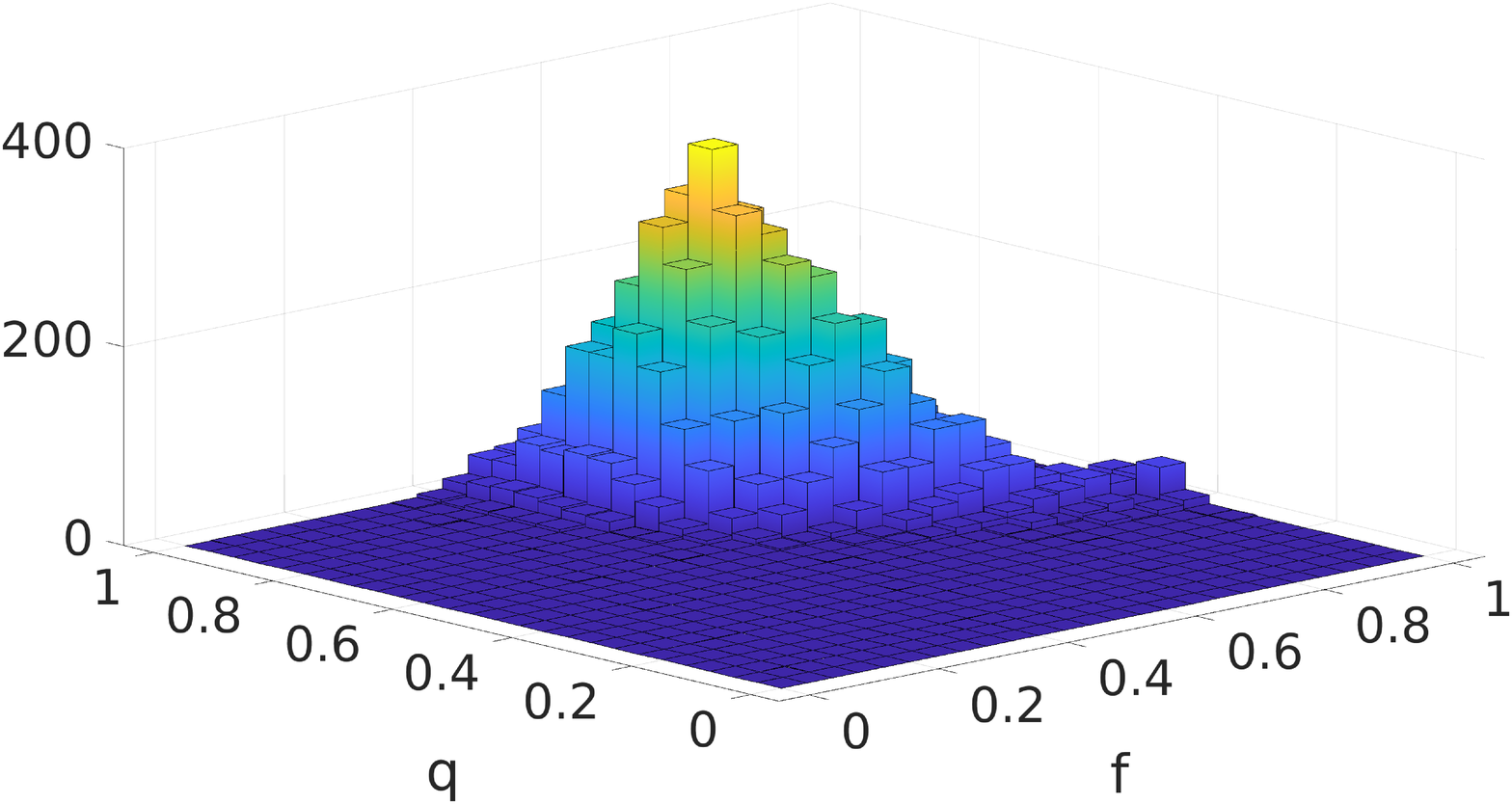}
        \includegraphics[width=0.32\textwidth]{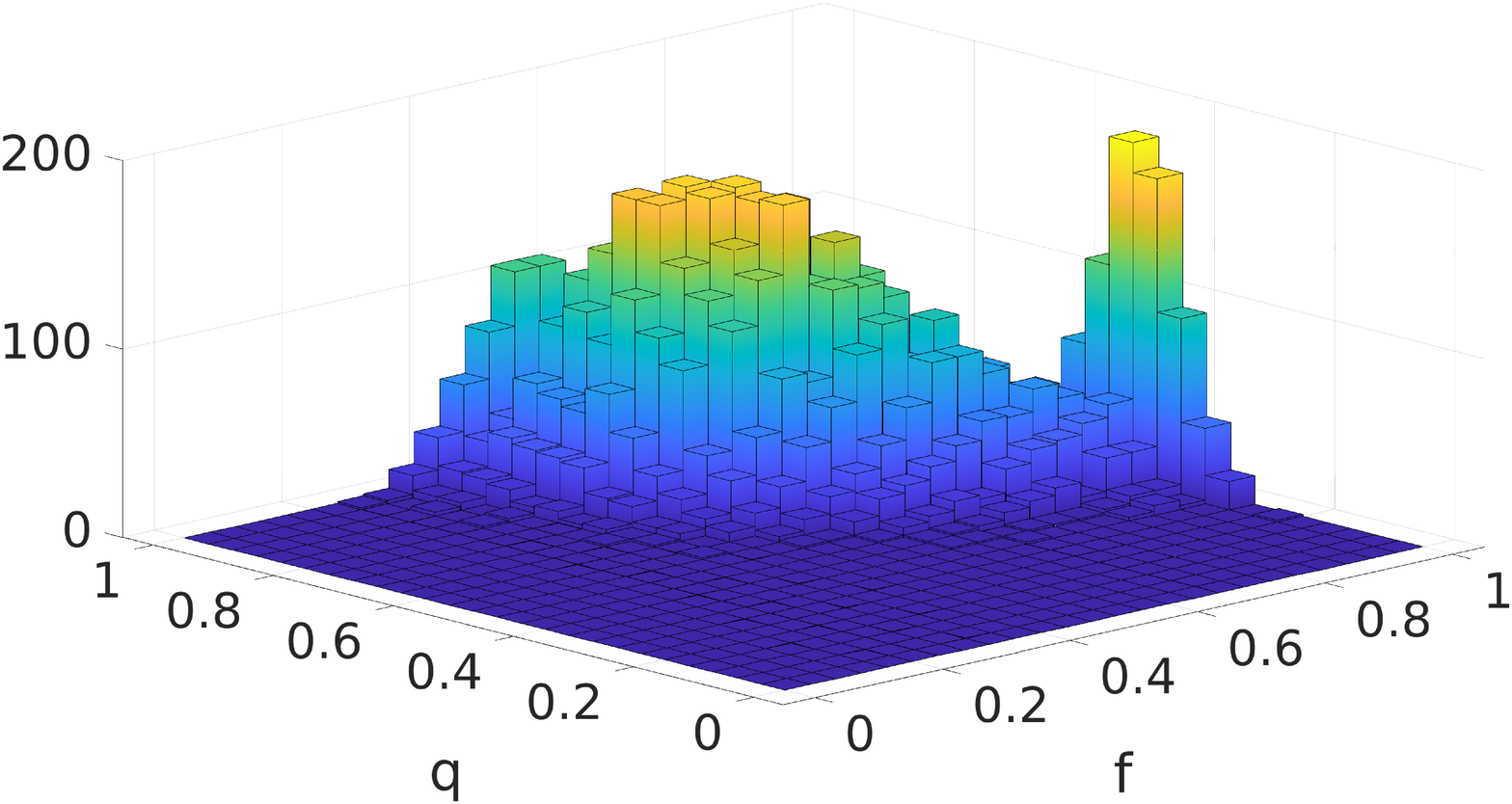}
        \includegraphics[width=0.32\textwidth]{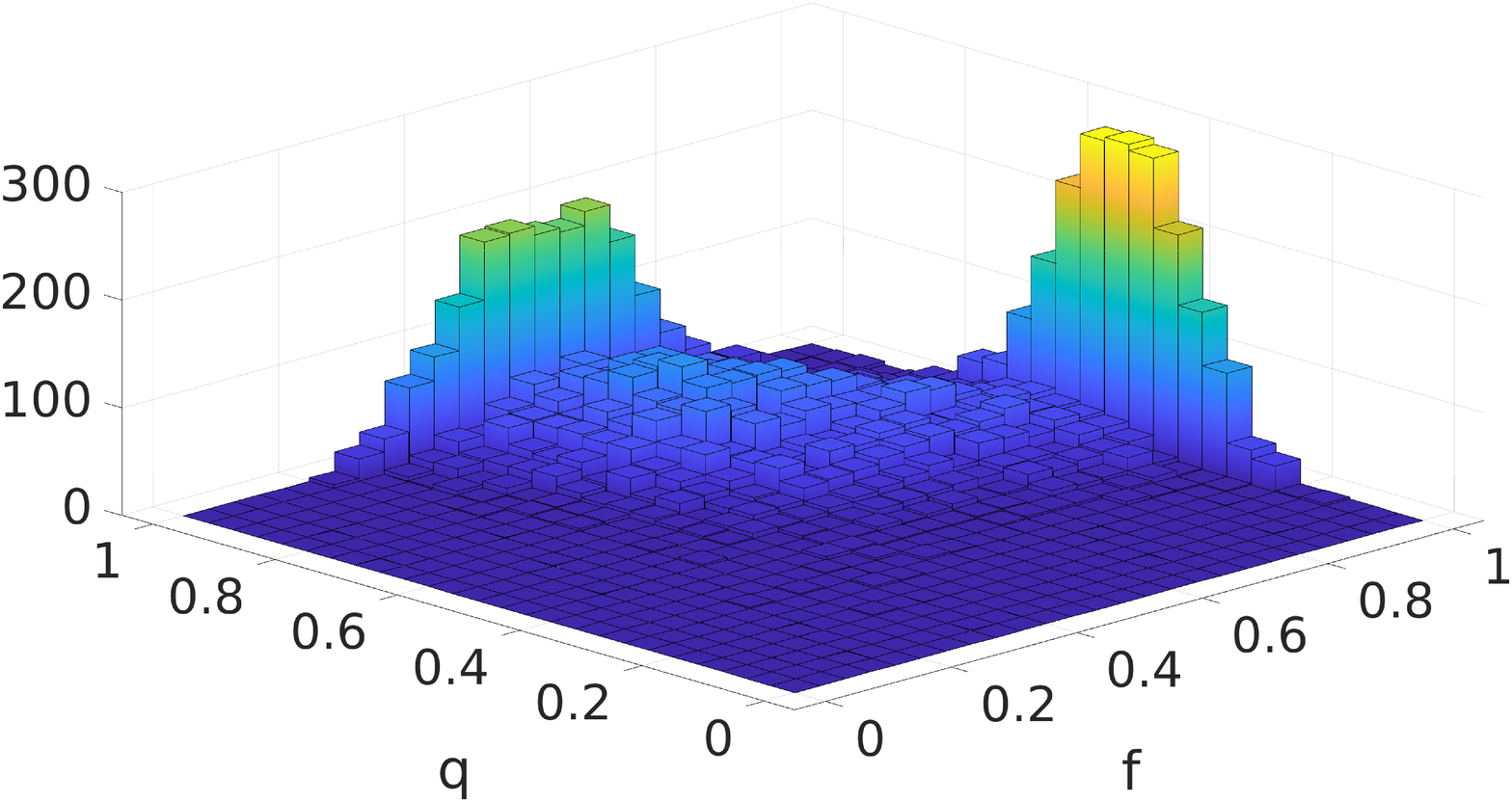}
    \caption{Histogram of the joint distribution of fitted values of the parameters ($q,f$) of ensembles of 10000 $\beta$-Hermite matrices with $f=0.7$ and $\beta=0.6$. The matrix sizes are from left to right $N=1000$, $N=500$, and $N=200$.}
    \label{fig:disp-interm}
\end{figure}

Now let us analyze in more detail a case of more practical interest with intermediate statistics between the two extremes. Let us think of a experimental spectrum with a $30\%$ of missing levels and dynamics between chaos and regularity with $\beta=0.6$, which we can represent here by one individual member of a $\beta$-ensemble generated with $f=0.7$ and $\beta=0.6$.
We can perform a fit to the two-parameter formula to obtain a first estimation of the parameters $q$ and $f$. But in view of the previous figures \ref{fig:disp-low} and \ref{fig:disp-high} it seems clear that the best estimation of the uncertainty of the parameters deals with a simulation of an ensemble of spectra of the same dimension of the experimental one to analyze the dispersion of values of the parameters more than with any error estimation from the fit.
We represent the results for this ensemble, for $N=1000$, in the left panel of Fig. \ref{fig:disp-interm} and we can see that the spread of values is wider than the one for higher values shown in Fig. \ref{fig:disp-high}, and one could have a first estimation of the uncertainties of the parameters from the widths of this distribution. Moreover, for lower dimensions one obtains even wider distributions of the parameters, as shown in the central panel of Fig. \ref{fig:disp-interm} ($N=500$) and the right panel of Fig. \ref{fig:disp-interm} ($N=200$). That is, the probability of obtaining the correct values of the parameters from a single fit decreases, as expected, confirming that a safe estimation of the parameters and their errors cannot be obtained from a single fit but analyzing the distributions of several simulations of ensembles. In these two panels we can also observe the same effect as in Fig. \ref{fig:disp-low} for the extreme of very uncorrelated spectra when the $P(s)$ can be well-fitted in some realizations of the ensemble with a combination of $f$ and $q$ such that one of them is close to unity and the other one is representative of the actual intermediate behavior of the NNSD. That is, the conclusion from the fit in these realizations would be that the intermediate statistics is due to just one reason, missing levels or intermediate dynamics.

Thus, in view of these examples, our conclusion and our recommendation when analyzing a spectrum of interest would be not only performing the fit but simulating several $\beta$-ensembles of the same dimension and observe the histograms of the distribution of the fitted results in order to try to obtain the most reliable estimation of the parameters and their uncertainties. For example, when having a spectrum like the ones in the left panel of Fig. \ref{fig:disp-interm} one could most probably obtain parameters near the center of the distribution and then start simulating ensembles with similar values of $f$ and $\beta$ and estimating their uncertainties from the width of the distributions. On the other hand, for a spectrum of lower dimension like the ones in the right panel of Fig. \ref{fig:disp-interm} one could obtain parameters around the right centered peak or parameters nearer the two extreme peaks around $f\simeq 1$ or $q \simeq 1$. Then for lower dimensions, when obtaining a value close to unity for any of the two parameters one should take this result with more caution and try to perform more simulations and checks to be sure of its reliability and make the best error estimation.

Apart from intrinsic limitations from low values of the parameters or low dimension, here we only have simulated spectra and have assumed that not any previous information on them is known, but in practical cases if one has some previous information about the type of dynamics of the system or the estimated fraction of missing levels, the limitations become less. In order to have the safest estimations of the uncertainties of the parameters we recommend to perform simulations of several ensembles in any case, but if the values can be restricted to some ranges from previous known information, much narrower distributions will be obtained with only one peak instead of two or three, even for low dimensions.

\section{Conclusions}


Quantum spectra with statistical properties intermediate between the GOE result of RMT and the Poisson uncorrelated spectra can be due to genuine intermediate properties between chaos and integrability or because there are missing levels destroying correlations in the level fluctuations induced by chaotic properties. But what happens when both things occur at the same time? Is it possible to distinguish the different causes of intermediate behavior and to quantify them? 

In this work, we have tried to answer this question. We have developed a two-parameter formula for fitting short-range spectral correlations with one of the parameters based on the Brody distribution accounting for the chaoticity of the spectra, $q$, and the other for the fraction of observed levels, $f$. The formulas work perfectly well when comparing with RMT results of the $\beta-$Hermite ensemble. 
This proofs the correctness of the formula for describing fractional level repulsion and missing levels in the spectra at the same time.

The theoretical interest of the formula is clear, as a first step to start to unravel the puzzle of the actual origin of intermediate behavior of fluctuations in a spectrum. Now, the next step would be to judge its applicability for actual experimental individual spectra. In this work we have shown some examples by simulating individual spectra from matrices of $\beta$-Hermite ensembles ($q=\beta$) and analyzed the accuracy and precision of the fit by computing the full distribution of the fitted results of the whole ensemble. From these distributions one can evaluate the probability to obtain reliable results depending on the values of $\beta$ and $f$ and the size of the spectrum, and estimate the uncertainties of the parameters. So this is our recommendation on how to proceed when using the two-parameter formula.

In view of the distributions of fitted results, we conclude that reliable results of both parameters at the same time can be obtained when the fraction of missing levels is not very large $f>0.6$ and the total number of levels in the analyzed sequence is large enough $N>1000$, assuming no previous information on the spectrum of interest. Though, in each particular case one can have some previous information on the values of $q$ and $f$ of the spectrum, which can help to shorten their ranges and improve the results even for lower dimensions. The formula would certainly be applicable to estimate the fraction of missing levels if we know in advance what is the value of the repulsion parameter $q$ or viceversa.

A similar procedure as the one we describe here for obtaining the two-parameter formula could be implemented for the Izrailev distribution, as it is also suitable to describe fractional level repulsion. We expect very similar results as the ones we present here. It should also be possible to start with the Berry-Robnik distribution that is able to fit the spectral statistics in the so-called 'far-semiclassical regime' which sometimes show a different behavior for small spacings, $P(0) \neq 0$ and its value is given by the fraction of the classical regular phase space. Thus, in these systems there is not fractional level repulsion but full level repulsion only for a fraction of levels of the spectrum. In order to obtain a suitable
two-parameter formula for the $P(s)$ in this case one should consider the Berry-Robnik distribution as a starting point \cite{Berry1984,Prosen1994}.

We have taken advantage of the $\beta$-Hermite ensemble for testing our formula as it is a simple way to scan the transition chaos ($\beta=1 )-$regularity ($\beta=0$) and its simple form results in an unrivalled speed-up of numerical simulations, but actual experimental spectra can be different from the $\beta$-ensemble, as the transition chaos-regularity is not universal. Thus, we make available this two-parameter formula and await results from experimental spectra to evaluate its real applicability and usefulness. The difficulties we have found in the estimation, specifically when the number of levels is limited, point also to the fact that, whenever possible, one should analyze together all the information available including short and long-range statistics and statistics of the widths in order to have more reliable conclusions regarding the chaoticity and the completeness of experimental spectra. 


{\it Acknowledgments:} This research has been supported by CSIC Research Platform on Quantum Technologies PTI-001.
We also acknowledge financial support from
European Union’s Horizon 2020 FET-Open project AVaQus
Grant No. 899561, Projects No. PGC2018-094180-B-I00, PID2019-106820RBC21, RTI2018-098868-B-I00 and PID2021-126998OB-I00
funded
by MCIN/AEI/10.13039/501100011033 and FEDER ”A way of making Europe”.


{\it Bibliography:}

\bibliography{reference2}

\end{document}